\documentclass[12pt]{article}

\begin{document}
\begin{center}
{\bf DIRAC'S QUANTIZATION OF MAXWELL'S THEORY ON NON-COMMUTATIVE SPACES}\\
\vspace{5mm}
 S.I. Kruglov \\
\vspace{5mm}
\textit{International Educational Centre, 2727 Steeles Ave. W, \# 202, \\
Toronto, Ontario, Canada M3J 3G9}
\end{center}

\begin{abstract}
Dirac's quantization of the Maxwell theory on non-commutative
spaces has been considered. First class constraints were found
which are the same as in classical electrodynamics. The gauge
covariant quantization of the non-linear equations of
electromagnetic fields on non-commutative spaces were studied. We
have found the extended Hamiltonian which leads to equations of
motion in the most general gauge covariant form. As a special
case, the gauge fixing approach on the basis of Dirac's brackets
has been investigated. The problem of the construction of the wave
function and physical observables have been discussed.
\end{abstract}

\section {Introduction}

Quantum field theories on non-commutative (NC) spaces are usually
formulated in terms of star-products of ordinary functions [1].
Non-commutative gauge theories are of interest now as they appear
in the superstring theory [2]. So, in the presence of the external
background magnetic field, NC coordinates can be introduced
naturally [2]. The NC gauge theories can be represented as
ordinary gauge theories (effective commutative theories) with the
same degrees of freedom, and with the additional deformation
parameter $\theta$ [3-5]. The Seiberg-Witten map between field
theory on NC spaces and the corresponding commutative field theory
allows us to formulate a Lagrange theory in terms of ordinary
fields. The Lagrangian of the corresponding action is expanded as
series of ordinary fields and a parameter that characterizes the
non-commutativity. This deformation parameter of the
non-commutative geometry, $\theta$, plays the role of coupling
constant of an effective Lagrangian. The parameter that
characterizes non-commutativity enters the coordinate commutation
relation [6,7]: $\left[\widehat{x}_\mu,\widehat{x}_\nu
\right]=i\theta_{\mu\nu}$.

In the present paper, we will apply the Dirac's quantization to
the Maxwell theory on non-commutative spaces. The corresponding
action in terms of ordinary fields and linear in the deformation
parameter has been derived in [5,8]. The one-loop corrections of
the vacuum polarization of photons and BRST-shift symmetry have
been considered in [9]. Some physical effects of propagation of
photons in such $\theta$-deformed Maxwell theory have been
investigated in [10-12]. The energy-momentum tensor and its trace
anomaly have been derived in [13].

The paper is organized as follows: In Sec. 2 the general Dirac's
procedure of quantization of $\theta$-deformed Maxwell theory is
considered. The extended Hamiltonian leading to gauge covariant
equations of motion is derived. The gauge fixing approach on the
basis of Dirac's brackets is studied in Sec. 3. The wave function
and physical observables are considered in Sec. 4. Section 5 is
devoted to the discussion.

We use Lorentz-Heaviside units, and set $\hbar=c=1$.

\section {The Canonical Hamiltonian and Equations of Motion}

The Maxwell Lagrangian density on NC spaces in terms of ordinary
fields in the order of ${\cal O}(\theta^2)$ is given by [8]
\begin{equation}
 {\cal L}=-\frac14F^2_{\mu \nu}+\frac18\theta_{\alpha\beta}
 F_{\alpha\beta}F^2_{\mu \nu}-\frac12\theta_{\alpha\beta}
 F_{\mu \alpha}F_{\nu \beta}F_{\mu \nu}+{\cal O}(\theta^2) ,
\label{1}
\end{equation}
where the field strength tensor is
 \begin{equation}
 F_{\mu \nu }=\partial _\mu A_\nu -\partial _\nu A_\mu ,
\label{2}
\end{equation}
$A_\mu =({\bf A},iA_0)$ is the vector-potential of the
electromagnetic field, $E_i =iF_{i4}$, $B_{i}
=\epsilon_{ijk}F_{jk}$ ($\epsilon_{123}=1$) are the electric field
and the magnetic induction field, respectively. The Lagrangian (1)
can also be represented as [13]
\begin{equation}
 {\cal L}=\frac12 \left( {\bf E}^2-{\bf B}^2 \right)\left[1+({\bf
 \theta}\cdot{\bf B})\right]-\left({\bf \theta}\cdot{\bf E}\right)\left({\bf
 E}\cdot{\bf B}\right)
 +{\cal O}(\theta^2) ,
\label{3}
\end{equation}
where $\theta_i=(1/2)\epsilon_{ijk}\theta_{jk}$, $\theta_{i4}=0$.
Here we use space-like components of the tensor $\theta_{\mu\nu}$
characterizing the non-commutativity because only in this case the
field theory possesses unitarity [14,15]. The terms in Eq. (3)
containing the non-commutative parameter $\theta$ violate CP -
symmetry. Moreover, particles in field theories on NC spaces have
the dipole moments violating the CP - symmetry but CPT - symmetry
remains unbroken [16,17]. It was verified that quantum
electrodynamics on NC spaces at one-loop level is a renormalizable
[18.8] and asymptotic free theory [19].

The Euler-Lagrange equations (field equations) follow from Eq.
(1), and are given by [10,11]
\begin{equation}
\frac{\partial}{\partial t}{\bf D}-\mbox{rot} {\bf H}=0
,\hspace{0.3in} \mbox{div}{\bf D}=0 , \label{4}
\end{equation}
where $(\mbox{rot} {\bf H})_{i} =\epsilon_{ijk}\partial_{j}H_{k}$
and $\mbox{div}{\bf D}=\partial_{i}D_{i}$. The displacement (${\bf
D}$) and magnetic (${\bf H}$) fields are defined as
\[
 {\bf D}={\bf E}+{\bf d},\hspace{0.3in}{\bf d}=({\bf \theta}\cdot
 {\bf B}){\bf E}-({\bf \theta}\cdot {\bf E}){\bf B}-
 ({\bf E}\cdot {\bf B}){\bf \theta} ,
\]
\vspace{-8mm}
\begin{equation}
\label{5}
\end{equation}
\vspace{-8mm}
\[
 {\bf H}={\bf B}+{\bf h},\hspace{0.3in}{\bf h}=({\bf \theta}\cdot
 {\bf B}){\bf B}+({\bf \theta}\cdot {\bf E}){\bf E}-
 \frac{1}{2}\left({\bf E}^2- {\bf B}^2\right){\bf \theta} .
\]
The second pair of equations, which is the consequence of Eq. (2),
is
\begin{equation}
 \partial_\mu \widetilde{F}_{\mu \nu }=0 ,
\label{6}
\end{equation}
where the dual tensor being
$\widetilde{F}_{\mu\nu}=(1/2)\varepsilon _{\mu \nu \alpha \beta }F
_{\alpha \beta}$, $\varepsilon _{\mu \nu \alpha \beta }$ is an
antisymmetric tensor Levy-Civita ($\varepsilon _{1234}=-i$). Eq.
(6) takes the form
\begin{equation}
\frac {\partial}{\partial t}{\bf B}+\mbox{rot}{\bf
E}=0,\hspace{0.3in} \mbox{div}{\bf B}=0 .
\label{7}
\end{equation}

Now we apply  Dirac's procedure [20] of gauge covariant
quantization to the Lagrangian (1) which leads to non-linear field
equations. The Lagrangian (1) is gauge invariant with the simple
gauge group $U(1)$ as well as ordinary electrodynamics. There is
here an infinite dimensional phase space because we deal with a
field theory. As usual the Lagrangian, $L$, and the action, $S$,
corresponding to the density of Maxwell's Lagrangian (1) are given
by
\begin{equation}
 L=\int d^3 x {\cal L} , \hspace{0.3in} S=\int dt L ,
\label{8}
\end{equation}
where $x_i$ (i=1,2,3) are spatial coordinates and $t$ is the time.
We study the time evolution of fields and come, therefore, to
``non-relativistic" formalism although the theory remains Lorentz
covariant. The potentials, $A_\mu$, in this formalism are
``coordinates" and ``velocities" are $\partial_0
A_\mu\equiv\partial A_\mu /\partial t $. According to the general
formalism [20], we find from Eq. (1), with the accuracy of ${\cal
O}(\theta^2)$, the following momenta:
\[
 \pi_i=\frac{\partial {\cal L}}{\partial(\partial_0 A_i)}=-E_i\left[1+({\bf
 \theta}\cdot{\bf B})\right]+\left({\bf \theta}\cdot{\bf E}\right)
  B_i+\left({\bf E}\cdot{\bf B}\right)\theta_i  ,
\]
\vspace{-8mm}
\begin{equation}
\label{9}
\end{equation}
\vspace{-8mm}
\[
\pi_0=\frac{\partial {\cal L}}{\partial (\partial_0 A_0)}=0 .
\]
The second equation in (9) gives a primary constraint
\begin{equation}
 \varphi_1 (x)=\pi_0 ,\hspace{0.3in}\varphi_1 (x)\approx 0 ,
\label{10}
\end{equation}
where we use Dirac's notation [20] $\approx$ for equations which
hold only weakly, i.e. $\varphi_1 (x)$ can have nonvanishing
Poisson brackets with some variables. Eq. (10) is an infinite set
of constraints for every space coordinate ${\bf x}$. From Eqs.
(9),(5), we come to the equality $\pi_i=-D_i$, i.e. the momentum
and the displacement field equal (with the opposite sign). Then,
using the known Poisson bracket $\{.,.\}$ between coordinates $A_i
(x)$ and momentum $\pi_i$, we arrive at
\begin{equation}
 \{A_i (\textbf{x},t),D_j(\textbf{y},t)\}=-\delta_{ij}
 \delta(\textbf{x}-\textbf{y}) .
\label{11}
\end{equation}
From Eq. (11) it is easy to find the Poisson bracket between the
magnetic induction field $B_i$ and the displacement field $D_j$:
\begin{equation}
 \{B_i (\textbf{x},t),D_j(\textbf{y},t)\}=\epsilon_{ijk}\partial_k
 \delta(\textbf{x}-\textbf{y})
 .
\label{12}
\end{equation}
The same relation holds in the Born-Infeld theory [20]. In the
quantized theory we have to make the substitution
\begin{equation}
 \{B,D\}\rightarrow -i\left[B,D\right] ,
\label{13}
\end{equation}
where $\left[B,D\right]=BD-DB$ is the quantum commutator. The
density of the Hamiltonian found from the relation ${\cal
H}=\pi_\mu
\partial_0 A_\mu-{\cal L}$, with the help of Eqs. (3), (9), is
given by
\begin{equation}
 {\cal H}=\frac12 \left( {\bf E}^2+{\bf B}^2 \right)\left[1+({\bf
 \theta}\cdot{\bf B})\right]-\left({\bf \theta}\cdot{\bf E}\right)\left({\bf
 E}\cdot{\bf B}\right)-\pi_m \partial_m A_0 .
\label{14}
\end{equation}
The primary constraint (10) should be a constant of motion and,
therefore, we have the condition
\begin{equation}
 \partial_0 \pi_0 =\{\pi_0,H\}=-\partial_m \pi_m= 0 ,
\label{15}
\end{equation}
where
\begin{equation}
 H=\int d^3x {\cal H} ,
\label{16}
\end{equation}
is the Hamiltonian. Eq. (15) guarantees that the primary
constraint (10) is conserved. The secondary constraint, found from
Eq. (15), is
\begin{equation}
 \varphi_2 (x)=\partial_m \pi_m ,\hspace{0.3in}\varphi_2 (x)\approx 0 .
\label{17}
\end{equation}
It should be noted that weak equalities, $\approx$, are not
compatible with the Poisson brackets [20].  Using the equality
$\pi_m=-D_m$, it is easy to see that the secondary constraint (17)
is simply the Gauss law (see Eq. (4)). The time evolution of the
secondary constraint
\begin{equation}
 \partial_0 \varphi_2 =\{\varphi_2,H\}\equiv 0 ,
\label{18}
\end{equation}
shows that there is no additional constraints. The Poisson bracket
between primary, $\varphi_1$, and secondary, $\varphi_2$,
constraints vanishes, $\{\varphi_1,\varphi_2\}=0$. Thus, all
constraints here are first class, and there are no second class
constrains, as in classical electrodynamics [20]. The primary and
secondary constraints can be considered on the same footing [20].
According to the general method [20], to acquire the total density
of Hamiltonian, we add to Eq. (14) Lagrange multiplier terms
$v(x)\pi_0$, $u(x)\partial_m \pi_m$, where $v(x)$ and $u(x)$ are
auxiliary variables which have no physical meaning, and are
connected with gauge degrees of freedom. As a result, we arrive at
the total density of Hamiltonian of Maxwell's theory on NC spaces:
\[
 {\cal H}_T=\frac12 \left( {\bf E}^2+{\bf B}^2 \right)\left[1+({\bf
 \theta}\cdot{\bf B})\right]-\left({\bf \theta}\cdot{\bf E}\right)\left({\bf
 E}\cdot{\bf B}\right)
\]
\vspace{-8mm}
\begin{equation}
\label{19}
\end{equation}
\vspace{-8mm}
\[
 -\pi_m \partial_m A_0+v(x)\pi_0+u(x)\partial_m \pi_m .
\]
The first class constraints in Eq. (19) generate gauge
transformations and Eq. (19) gives the set of Hamiltonians. As the
physical space is the constraint surface, we get the energy from
the Hamiltonian on the constraint surface. Thus, the density
energy, found from Eq. (19), is given by
\begin{equation}
 {\cal E}=\frac12 \left( {\bf E}^2+{\bf B}^2 \right)\left[1+({\bf
 \theta}\cdot{\bf B})\right]-\left({\bf \theta}\cdot{\bf E}\right)\left({\bf
 E}\cdot{\bf B}\right).
\label{20}
\end{equation}
The same expression, Eq. (20), was obtained in [13] from another
procedure. To obtain equations of motion we have to express the
density of Hamiltonian (19) in terms of fields, $A_\mu$, and
momenta, $\pi_\mu$. For this, using Eq. (9), we find, with the
accuracy of ${\cal O}(\theta^2)$, the electric field
\begin{equation}
E_i=-\pi_i\left[1-({\bf
 \theta}\cdot{\bf B})\right]-\left({\bf \theta}\cdot{\bf \pi}\right)
  B_i-\left({\bf \pi}\cdot{\bf B}\right)\theta_i  .
\label{21}
\end{equation}
With the help of Eq. (21) and the equality $D_i=-\pi_i$, the total
density of Hamiltonian (19) takes the form
\[
 {\cal H}_T=\frac {{\bf \pi}^2+{\bf B}^2}{2}+({\bf
 \theta}\cdot{\bf B})\frac {{\bf B}^2-{\bf \pi}^2}{2}
 +\left({\bf \theta}\cdot{\bf \pi}\right)\left({\bf
 \pi}\cdot{\bf B}\right)
\]
\vspace{-8mm}
\begin{equation}
\label{22}
\end{equation}
\vspace{-8mm}
\[
 +v(x)\pi_0+\left(u(x)+A_0\right)
 \partial_m \pi_m ,
\]
where ${\bf B}=\mbox{rot}{\bf A}$ and fields $A_i$ enter the
density of Hamiltonian (22) in the form of the $\mbox{rot}{\bf
A}$. We took into account Eq. (16) and an integration by parts to
get the term $A_0\partial_m \pi_m $ in Eq. (22). As the function
$u(x)$ is arbitrary, we can make the substitution $u'(x)=u(x)+A_0$
(see [20]) in Eq. (22), so that $u'(x)$ will also be arbitrary,
and the component $A_0$ is absorbed by the function $u(x)$. After
this substitution, the $A_0$ does not enter the Hamiltonian and
its dynamics is not defined by the Hamiltonian. This means that
the component $A_0$ is not the physical degree of freedom. The
same concerns the component $\pi_0$, which is zero (see (10)). The
role of terms $v(x)\pi_0+u(x)\partial_m \pi_m$ in Eq. (22) is to
generate gauge transformations of fields which do not affect the
physical state of the system.

The total density of Hamiltonian allows us to obtain the time
evolution of fields. With the help of the Hamiltonian equations we
find
\[
\partial_0 A_i=\{A_i,H\}=\frac{\delta H}{\delta \pi_i}
\]
\vspace{-8mm}
\begin{equation}
\label{23}
\end{equation}
\vspace{-8mm}
\[
=\pi_i\left[1-({\bf \theta}\cdot{\bf B})\right]+\left({\bf
B}\cdot{\bf \pi}\right) \theta_i+\left({\bf \pi}\cdot{\bf
\theta}\right)B_i -\partial_i A_0-\partial_i u(x) ,
\]
\[
\partial_0 \pi_i=\{\pi_i, H\}=-\frac{\delta H}{\delta A_i}
=\partial_n\left\{\left[ (\partial_n A_i)-(\partial_i A_n)\right]
\left[1+({\bf
 \theta}\cdot{\bf B})\right]\right\}
\]
\vspace{-8mm}
\begin{equation}
\label{24}
\end{equation}
\vspace{-8mm}
\[
 +\varepsilon_{iab}\partial_b\left[
\theta_a\frac {{\bf B}^2-{\bf \pi}^2}{2}+\pi_a ({\bf \pi}\cdot{\bf
\theta})\right] ,
\]
\begin{equation}
\partial_0 A_0=\{A_0,H\}=\frac{\delta H}{\delta \pi_0}=v(x),\hspace{0.1in}
\partial_0 \pi_0=\{\pi_0,H\}=-\frac{\delta H}{\delta
A_0}=-\partial_m \pi_m . \label{25}
\end{equation}
Eq. (24) coincides with the first equation in (4) taking into
consideration the definition (5), and Eq. (23) is nothing but the
gauge covariant form of Eq. (21). The second equation in (4),
Gauss's law, is the secondary constraint in the Hamiltonian
formalism. As a particular case at $v(x)=\partial_0 u'(x)$
($u'(x)=u(x)+A_0$), we arrive from Eqs. (23), (25) at the
relativistic form of ordinary gauge transformations generated by
constraints:
\begin{equation}
A'_\mu (x)=A_\mu (x)+\partial_\mu \Lambda(x) , \label{26}
\end{equation}
where $\Lambda (x)=\int dt u'(x)$. But in general case, there are
here two arbitrary functions, $v(x)$, $u(x)$. Thus, the
Hamiltonian equations (23), (24) give the time evolution of
physical fields which are gauge-equivalent to some solutions of
the Euler-Lagrange equations. In this approach, the first class
constraints initiate gauge transformations and realize a gauge
algebra representation. Eq. (25) represents the time evolution of
non-physical fields. According to Eq. (25) the component $A_0$ is
arbitrary function connecting with gauge degree of freedom, and
$\pi_0$, $\partial_m \pi_m$ equal zero as constraints.

\section {The Coulomb Gauge and Quantization of Second-Class Constraints}

Let us consider the Coulomb (radiation) gauge constraints using
the gauge freedom of the $\theta$-deformed Maxwell theory. For
classical electrodynamics such a procedure was considered in [21]
(see also [22], [23]). It should be noted that the gauge fixing
procedure is beyond the Dirac's approach. The gauge degrees of
freedom are present in the Dirac method. So, in this section, we
consider gauge fixing approach.

 With the help of the gauge freedom, described by two functions,
 $v(x)$, $u(x)$, or Eq. (26), we can impose new constraints as
 follows:
\begin{equation}
\varphi_3 (x)=A_0\approx 0 ,\hspace{0.3in}\varphi_4 (x)=\partial_m
A_m\approx 0 . \label{27}
\end{equation}
The Coulomb gauge (27) does not violate the equations of motion.
After fixing two components of vector-potential $A_\mu$ in
accordance with Eqs. (27), the first class constraints (10), (17)
become second class constraints. Indeed, the non-zero Poisson
brackets of functions $\varphi_1$, Eq. (10), $\varphi_2$, Eq. (17)
and $\varphi_3$, $\varphi_4$, Eq. (27), are (see [21])
\begin{equation}
\{\varphi_1 (\textbf{x},t),\varphi_3
(\textbf{y},t)\}=-\delta(\textbf{x}-\textbf{y}) ,\hspace{0.1in}
\{\varphi_2 (\textbf{x},t),\varphi_4
(\textbf{y},t)\}=\Delta_x\delta(\textbf{x}-\textbf{y}) ,
\label{28}
\end{equation}
where $\Delta_x\equiv\partial^2/(\partial x_m)^2$. The pairs of
``coordinates" $Q_i$ and conjugated momenta $P_i$ ($i=1,2$) are
defined as follows (see [23]):
\begin{equation}
Q_i=(A_0,\partial_m
A_m),\hspace{0.3in}P_i=(\pi_0,-\Delta^{-1}_x\partial_m \pi_m) ,
\label{29}
\end{equation}
and equations
\begin{equation}
\{Q_i(\textbf{x},t),P_j(\textbf{y},t)\}=\delta_{ij}
\delta(\textbf{x}-\textbf{y}) ,\label{30}
\end{equation}
\begin{equation}
\Delta^{-1}_x=-\frac{1}{4\pi|\textbf{x}|} ,\hspace{0.3in}\Delta_x
\frac{1}{4\pi|\textbf{x}|}=-\delta(\textbf{x}) ,\label{31}
\end{equation}
hold. Pairs of canonical variables $Q_i,P_i$, Eq. (29), do not
describe a true physical degrees of freedom. Therefore, in the
quantum theory these operators must be eliminated. Defining the
matrix of Poisson brackets as [21]
\begin{equation}
C_{ij}=\{\varphi_i (\textbf{x},t),\varphi_j (\textbf{y},t)\} ,
\label{32}
\end{equation}
so that the inverse matrix $C_{ij}^{-1}$ exists [21], we may
introduce the Dirac bracket [20], [21]:
\[
\{A(\textbf{x},t),B(\textbf{y},t)\}^*=\{A(\textbf{x},t),B(\textbf{y},t)\}
\]
\vspace{-8mm}
\begin{equation}
\label{33}
\end{equation}
\vspace{-8mm}
\[
-\int d^3zd^3w\{A(\textbf{x},t),\varphi_\alpha(\textbf{z},t)\}
C^{-1}_{\alpha\beta}(z,w)\{\varphi_\beta(\textbf{w},t),
B(\textbf{y},t)\} .
\]
The inverse matrix $C_{ij}^{-1}$ obeys the equation
\begin{equation}
\int d^3z C_{\alpha\gamma}(x,z)C^{-1}_{\gamma\beta}(z,y)=
\delta_{\alpha\beta}\delta(\textbf{x}-\textbf{y}) , \label{34}
\end{equation}
and is given by
\begin{equation}
C^{-1}_{\alpha\beta}(x,y)=\left(
\begin{array}{cccc}
0 & 0 & \delta(\textbf{x}-\textbf{y}) & 0 \\
0 & 0 & 0 & \frac{1}{4\pi|\textbf{x}-\textbf{y}|} \\
-\delta(\textbf{x}-\textbf{y}) & 0 & 0 & 0 \\
0 & -\frac{1}{4\pi|\textbf{x}-\textbf{y}|} & 0 & 0
\end{array}\right) . \label{35}
\end{equation}
Using the definition of Dirac's bracket (33) and Eq. (35), and
imposing the boundary condition that the fields vanish at
infinity, we arrive at the same expression for the Dirac brackets
as in classical electrodynamics (see [21]):
\begin{equation}
\{\pi_0(\textbf{x},t),A_0(\textbf{y},t)\}^*=
\{\pi_0(\textbf{x},t),A_i(\textbf{y},t)\}^*=
\{\pi_i(\textbf{x},t),A_0(\textbf{y},t)\}^*=0 ,\label{36}
\end{equation}
\begin{equation}
\{\pi_i(\textbf{x},t),A_j(\textbf{y},t)\}^*=
-\delta_{ij}\delta(\textbf{x}-\textbf{y})+\frac{\partial^2}{\partial
x_i \partial y_j} \frac{1}{4\pi|\textbf{x}-\textbf{y}|}
\hspace{0.2in}(i,j=1,2,3) ,\label{37}
\end{equation}
\begin{equation}
\{\pi_\mu(\textbf{x},t),\pi_\nu(\textbf{y},t)\}^*=
\{A_\mu(\textbf{x},t),A_\nu(\textbf{y},t)\}^*=0
\hspace{0.3in}(\mu, \nu=1,2,3,4) .\label{38}
\end{equation}
Using the well-known Fourier transformation of the Coulomb
potential [24]
\begin{equation}
\int\frac{d^3 x}{4\pi|\textbf{x}|}e^{-i\textbf{k}\cdot\textbf{x}}=
\frac{1}{|\textbf{k}|^{2}} ,\label{39}
\end{equation}
Eq, (37) takes the form
\begin{equation}
\{\pi_i(\textbf{k}),A_j(\textbf{q})\}^*=-(2\pi)^3
\delta(\textbf{k+q})\left(\delta_{ij}-\frac{k_i
k_j}{\textbf{k}^2}\right)
 .\label{40}
\end{equation}
So, the projection operator in the right side of Eq. (40):
\[
\Pi=(\Pi_{ij})=\left(\delta_{ij}-\frac{k_i
k_j}{\textbf{k}^2}\right)
\]
with the properties $\Pi^2=\Pi$, $\Pi\textbf{k}=0$, extracts the
physical transverse components of vectors.

According to the definition (33), the Dirac bracket of any
operator $A$ with a second class constraint, $\varphi_\alpha$,
vanishes, $\{A,\varphi_\alpha\}^*=0$. Therefore, following the
prescription [20], we can set all second class constraints
strongly to zero. As a result, only two transverse components of
the vector potential $A_\mu$ and momentum $\pi_\mu$ are physical
independent variables. Thus, pairs of operators (29) are absent in
the reduced physical phase space. Then the physical Hamiltonian of
fully constrained $\theta$-deformed Maxwell's theory becomes
\begin{equation}
 H^{ph}=\int d^3 x {\cal E}=\int d^3 x \left\{\frac12 \left( {\bf E}^2+
 {\bf B}^2 \right)\left[1+({\bf \theta}\cdot{\bf B})\right]-
 \left({\bf \theta}\cdot{\bf E}\right)\left({\bf
 E}\cdot{\bf B}\right)\right\} .
\label{41}
\end{equation}
Equations of motion obtained from Eq. (41) are given by
\[
\{{\bf A},H^{ph}\}^*=\partial_0 {\bf A}
\]
\vspace{-8mm}
\begin{equation}
\label{42}
\end{equation}
\vspace{-8mm}
\[
={\bf \pi}\left[1-({\bf \theta}\cdot{\bf B})\right]+\left({\bf
B}\cdot{\bf \pi}\right) {\bf \theta}+\left({\bf \pi}\cdot{\bf
\theta}\right){\bf B} ,
\]
\[
\{{\bf \pi},H^{ph}\}^*=\partial_0 {\bf \pi}
\]
\vspace{-8mm}
\begin{equation}
\label{43}
\end{equation}
\vspace{-8mm}
\[
 =-\mbox{rot}\left\{{\bf B} \left[1+({\bf
 \theta}\cdot{\bf B})\right]+ {\bf \theta}\frac {{\bf B}^2-
 {\bf \pi}^2}{2}+{\bf \pi}({\bf \pi}\cdot{\bf
\theta})\right\} .
\]
In the quantum theory, with the presents of second class
constraints, we have to replace Dirac's bracket by the quantum
commutator according to the prescription
$\{.,.\}^*\rightarrow-i[.,.]$. Only transverse components of the
vector potential $A_\mu$ are physical degrees of freedom, and they
remain in the theory.

\section {Wave Function, Observables and Quantization of
First-Class Constraints}

We do not imply here the gauge conditions, described in the
previous section, and we follow the direct Dirac method. The gauge
degrees of freedom present as operators in a bigger linear space
and constraints are operators which act on Dirac's space. One
needs to extract the physical sub-space by imposing conditions.

Let us consider the problem of constructing physical states in the
Hilbert space and observables. In quantized theory, the dynamical
variables $\hat{A}_i$ and $\hat{\pi}_i=-\hat{D}_i$ obey here the
commutation relation (see (11)):
\begin{equation}
 [\hat{A}_i (\textbf{x},t),\hat{D}_j(\textbf{y},t)]=-i\delta_{ij}
 \delta(\textbf{x}-\textbf{y}) .
\label{44}
\end{equation}
The wave function $|\Psi \rangle $ satisfies the Schr\"{o}dinger
equation
\begin{equation}
i\frac{d|\Psi \rangle}{dt}=H|\Psi \rangle , \label{45}
\end{equation}
where $H$ is the Hamiltonian (16) with the density (19). In
accordance with [20] the wave function (the state) must obey the
following equations
\begin{equation}
\hat{D}_0 |\Psi \rangle=0 ,\hspace{0.3in}\partial_m \hat{D}_m|\Psi
\rangle=0 , \label{46}
\end{equation}
i.e., the physical state remains unchanged, and as a result, it is
invariant under the gauge transformations. In the coordinate
representation the operators of the ``coordinate" $\hat{A}_i$ and
the momentum $\hat{\pi}_i=-\hat{D}_i$ are given by
\begin{equation}
\hat{A}_i \Psi [A]=A_i \Psi [A],\hspace{0.3in} \hat{D}_\mu (x)\Psi
[A] =i\frac{\delta\Psi [A]}{\delta A_\mu (x)} ,\label{47}
\end{equation}
where $\Psi [A]$ is the wave functional (the vector of the state).
In this representation Eqs. (46) take the form [25]
\begin{equation}
\frac{\delta\Psi [A]}{\delta A_0 (x)}=0 ,\hspace{0.3in}\partial_i
\frac{\delta\Psi [A]}{\delta A_i (x)} =0 . \label{48}
\end{equation}
Thus, constraints are restrictions on the state $|\Psi \rangle$.
Both equations (46) (or (48)) are compatible if $[\hat{D}_0,
\partial_m \hat{D}_m]=0$; this is the case. So, constraints do not
change the physical states $|\Psi \rangle$ being gauge invariant
quantities, and are the generators of the gauge symmetry. The
method of gauge fixing, described in the previous section, leads
to the reduced phase space, and is equivalent to the Dirac
approach.

The fields $\textbf{E}$, $\textbf{B}$, $\textbf{D}$,, $\textbf{H}$
are invariants of the gauge transformations and are observables
(measurable quantities). As usual, real observables are
represented by the Hermitian operators and must not depend on
$A_0$.

To have normalized states and a physical interpretation to the
theory, one needs to construct a scalar product of wave functions
(functionals). Obviously, this product is given by the functional
integral. Then, however, we should take into account the gauge
degrees of freedom, and insert a gauge condition. As a result, we
arrive at the necessity to introduce ghosts. Such a procedure is
beyond the Dirac approach. The second way is to use the Fock basis
ignoring the wave functionals. The problem of constructing the
vacuum state and the scalar product on the physical state space
may be realized, with the help of the Fock representation, in the
same manner as in the case of classical electrodynamics (see
[25]). This procedure may involve, however, the introducing
negative norm states which should not appear in the physical
spectrum.

\section {Conclusion}

We have considered quantization of the Maxwell theory on
non-commutative spaces taking into consideration first class
constraints as well as introducing second class constraints and
the Dirac bracket. The procedure of Dirac's quantization here, on
the basis of first class constraints and the Poisson bracket, is
similar to the quantization of classical electrodynamics because
the gauge group is the same: $U(1)$. The difference is that field
equations are nonlinear in the case of the $\theta$-deformed
Maxwell theory, and, as a result, the quantization of a theory is
more complicated. Dirac's method of quantization has an advantage
compared to the reduced phase space approach (see [22]) that it
does not violate the Lorentz invariance and locality in space.

The quantization of Maxwell's theory on non-commutative spaces
within BRST-scheme, by the inclusion of the ghosts, was performed
in [8,9,26].

Let me now mention some interesting phenomenon impacting on NC
quantum electrodynamics (NCQED). NC theory can be verified at high
energy using $e^+e^-$ and hadronic collisions and low energy
precision experiments for measurements of anomalous magnetic (AM)
and electric dipole (ED) moments, as well as other CP-violating
effects [27]. So, $e^+e^-$, $e\gamma$, and $\gamma \gamma$  cross
sections within the NCQED framework depend on $\theta_{\mu\nu}$
[28]. In particular, for M\"{o}ller and Bhabha scattering, cross
sections depend on $\theta_{ij}$ (when i,j=1,2,3) and
$\theta_{0j}$, respectively. Note the Aharonov-Bohm effect for
high energy electrons might also influence the boundary on
$\theta$ [29].

There are constraints on $\theta$ parameters when splitting levels
of energy in positronium and the Lamb shift [30]. Electron ED
moments and muon AM moments give the boundary of $\theta<10^{-4}$
TeV$^{-1}$ [16,31], and $\theta<m_\mu ^{-1}$ [32], respectively.
In the electroweak sector, the CP violation provides the NC effect
$\theta<1$ TeV$^{-1}$ [31,33].

NC geometry contributes some effects in cosmology [34] because in
the early universe, at temperatures above $\theta ^{-1/2}$ NC
effects were important. That may explain the detection of high
energy photons in the cosmos [35] and a global structure after the
universe had expanded.


\begin{thebibliography}{999}

\bibitem{1} J.E. Moyal, Proc. Cambridge Phil. Soc. {\bf 45}, 99 (1949).
\bibitem{2} N. Seiberg and E. Witten, JHEP {\bf 9909}, 032 (1999).
\bibitem{3} J. Madore, S. Schraml, P. Schupp and J. Wess, Eur.
Phys. J. {\bf C 16}, 161 (2000); hep-th/0001203.
\bibitem{4} B. Jurco, S. Schraml, P. Schupp and J. Wess, Eur.
Phys. J. {\bf C 17}, 521 (2000); hep-th/0006246.
\bibitem{5} B. Jurco, L. Moller, S. Schraml, P. Schupp and J. Wess, Eur.
Phys. J. {\bf C 21}, 383 (2001); hep-th/0104153.
\bibitem{6} H. Snyder, Phys. Rev. {\bf 71}, 38 (1947); {\bf 72}, 68 (1947).
\bibitem{7} A. Connes, Noncommutative Geometry (Academic Press, 1994).
\bibitem{8} A. Bichl, J. Grimstrup, L. Popp, M. Schweda and
R. Wulkenhaar, hep-th/0102044.
\bibitem{9} I. Fruhwirth, J.M. Grimstrup, Z. Morsli, L. Popp and M. Schweda,
hep-th/0202092.
\bibitem{10} R. Jackiw, Nucl. Phys. Proc. Suppl. {\bf 108}, 30 (2002);
hep-th/0110057.
\bibitem{11} Z. Guralnic, R. Jackiw, S.Y. Pi and A.P.Polychronakos,
Phys. Lett. {\bf B 517}, 450 (2001); hep-th/0106044.
\bibitem{12} R.G. Cai, Phys. Lett. {\bf B 517}, 457 (2001); hep-th/0106047.
\bibitem{13} S.I. Kruglov, Annales Fond. Broglie (in press); hep-th/0110059.
\bibitem{14} J. Gomis and T. Mehen, Nucl. Phys. {\bf B591}, 265 (2000).
\bibitem{15} O. Aharony, J. Gomis and T. Mehen, JHEP {\bf 0009}, 023 (2000).
\bibitem{16} I.F. Riad and M.M. Sheikh-Jabbari, JHEP {\bf 0008}, 045
(2000).
\bibitem{17} M.M. Sheikh-Jabbari, Phys. Rev. Lett. {\bf 84}, 5265 (2000).
\bibitem{18} M. Hayakawa, Phys. Lett. {\bf B478}, 394 (2000).
\bibitem{19} C.P. Martin and D. Sanchez-Ruiz, Phys. Rev. Lett. {\bf 83},
476 (1999).
\bibitem{20} P.A.M. Dirac, Lectures on Quantum Mechanics (Yeshiva
University, New York, 1964).
\bibitem{21} A. Hanson, T. Regge, C. Teitelboim, Constrained
Hamiltonian Systems (Accademia Nationale Dei Lincei, Roma, 1976).
\bibitem{22} M. Henneaux and C. Teitelboim, Quantization of Gauge
Systems (Princeton University Press, Princeton, New Jesey, 1992).
\bibitem{23} D.M. Gitman and I.V. Tyutin, Quantization of Fields
with Constraints (Springer Series in Nuclear and Particle Physics,
Springer-Verlag, Berlin, Heidelberg, 1990).
\bibitem{24} J.D. Bjorken and S.D. Drell, Relativistic Quantum
Fields (McGraw-Hill, Ind., New York, 1964).
\bibitem{25} H.J. Matschull, quant-ph/9606031.
\bibitem{26} R. Wulkenhaar, JHEP {\bf 0203}, 024 (2002); hep-th/0112248.
\bibitem{27} I. Hinchliffe and N. Kersting, hep-ph/0205040.
\bibitem{28} J.L. Hewett, F.J. Petriello and T.G. Rizzo, Phys. Rev. {\bf D 64},
075012 (2001); P. Mathews, Phys. Rev. {\bf D 63}, 075007 (2001);
S.-W. Baek et al., Phys. Rev. {\bf D 64}, 056001 (2001).
\bibitem{29} H. Falomir et al., hep-th/0203260.
\bibitem{30} M. Chaichian, M.M. Sheikh-Jabbari and T.Tureanu, Phys. Rev.
Lett. {\bf 86}, 2716 (2001); M. Haghighat, S.M. Zebarjad and F.
Loran, hep-ph/0109105.
\bibitem{31} I. Hinchliffe and N. Kersting, Phys. Rev. {\bf D 64}, 116007
(2001).
\bibitem{32} X.-J. Wang and M.-L. Yan, JHEP {\bf 03}, 047 (2002).
\bibitem{33} Z. Chang and Z.-Z. Xing, hep-ph/0204255.
\bibitem{34} H. Garcia-Compean, O. Obregon and C. Ramirez, Phys. Rev. Lett. {\bf 88},
161301 (2002).
\bibitem{35} T. Tamaki and T. Harada, Phys. Rev. {\bf D 65}, 083003 (2002).

\end{thebibliography}
\end{document}